# NON LOCAL TRANSPORT IN PBTE WIDE PARABOLIC QUANTUM WELLS

J. OSWALD, G. HEIGL, G. SPAN, A. HOMER, P. GANITZER

*Institute of Physics, University of Leoben, Franz Josef Str. 18*
*A-8700 Leoben, Austria*

D.K. MAUDE, J.C. PORTAL

*High Magnetic Field Laboratory, CNRS, 25 Avenue des Martyrs*
*BP 166 Grenoble, France*

The results of non-local experiments in different contact configurations are discussed in terms of a non-local behaviour of the contact arms. It is shown that the observed reproducible fluctuations can be understood to result from fluctuations of a non-local bulk current in the contact arms. The fluctuations are explained by edge channel backscattering because of potential fluctuations in the bulk region.

## 1 Basic considerations about non-local transport

Basically the presence of edge channels (EC) makes it possible to have transport of current to a remote region. However, a potential difference in a remote region requires dissipation which leads automatically to a potential difference between opposite EC's of the remote path. As a result, a net edge current to the remote region appears, which finally requires an independent return path in order to maintain the current balance. The modelling of an EC-related non-local behaviour must therefore include also an alternative current path in addition to the EC's. This is schematically shown in the equivalent circuit of Fig.1. In the standard Quantum Hall (QH) systems such an alternative path occurs only in the regime between (QH) plateaus[1]. In this regime the states of the upper Landau level (LL) become extended into the bulk region and are almost de-coupled from the EC's of the lower LL's. In this way we have an independent bulk- and EC-system even in the standard single valley situation. However, well developed DOS-gaps between the LL's are required and therefore the observation of non-local EC-transport in standard single valley systems must be closely related to the appearance of the QHE. The presence of the bulk conduction in addition to edge channels leads to different potentials at the contacts (non-zero $R_{xx}$) which causes also a non-zero bulk current in the voltage probes according to the right part of Fig.1. Consequently this bulk current in the voltage probes leads to a non-local behaviour in the contact arms.

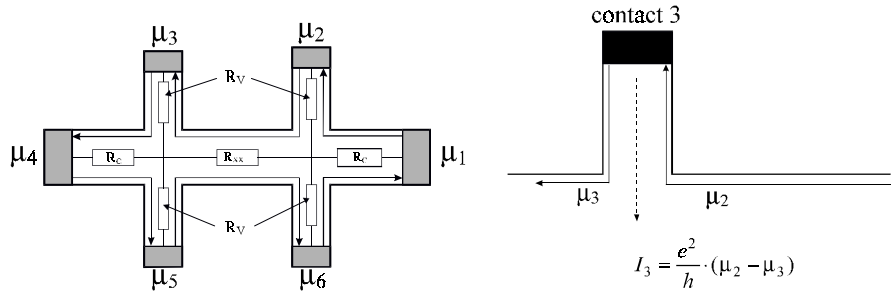

**Figure 1 left:** Equivalent circuit of a Hall bar sample combining EC- and bulk transport. The EC's are represented by arrows connecting directly the reservoirs of the contacts. The bulk system is represented by the resistor network. **Right:** Situation at contact 3. From the difference of potentials $\mu_2$ and $\mu_3$ a net edge current to the reservoir of contact 3 appears. This causes a bulk current (dashed arrow) leaving contact 3.

An influence of this non-local current on the total sample behaviour can result in two ways. (i) The non-local voltage drop in the contact arms adds directly to the local voltage between the voltage probes. (ii) The current in the contact arms leads to a global current redistribution between the EC-system and the bulk system of the sample which finally causes also a change of the local signals.

## 2  Wide quantum wells based on PbTe

In quasi-3D wide quantum wells the energy splitting between individual Landau subband levels will be insufficient to create an independent bulk- and EC-conduction within a single electron system. However, the presence of independent electron systems in different valleys of a many valley semiconductor like PbTe can also meet the requirements for non-local transport. One major advantage of PbTe is the possibility to achieve a high electron mobility ($\mu > 10^5$ cm$^2$V$^{-1}$s$^{-1}$) even without remote doping. The details about PbTe can be obtained from reference 2. The most important fact to mention here is that we get two sets of subbands and two sets of LL's[3]. For typical sample parameters we get a thickness of the conducting channel of about 330nm and an electron sheet density of about $10^{13}$cm$^{-2}$. This results in a subband splitting of the order of 1 meV for the small effective mass and about an order of magnitude smaller for the large effective mass. Consequently the small mass can lead already to EC-conduction while the large mass will create mainly a bulk system. In this way we can get a system which also fits to the equivalent circuit in Fig.1. The necessary suppression of intervalley scattering must be considered as an assumption and cannot be discussed within the scope of this paper. It is important to note that a subband splitting of about 1 meV is already of the order of the native potential fluctuations which result to be of the order of 1...2meV[4]. Therefore we can expect that a system of coupled magnetic bound states is created in the bulk region, which can provide an effective mechanism for EC-backscattering in the narrow contact arms of the potential probes.

Conductance fluctuations in PbTe WPQW's in local and non-local contact configuration have been reported for the first time in reference 3 and 5, where also further details about the experiments and the samples can be found. The fluctuation amplitude in the magneto conductance was found to be close to $e^2/h$ although the samples have been of macroscopic size (butterfly Hall bar of width: 100 μm, length between voltage probes: 200 μm, total length between the current contacts: 600μm). In this paper we concentrate on the results of non-local experiments which are shown in Fig.2.

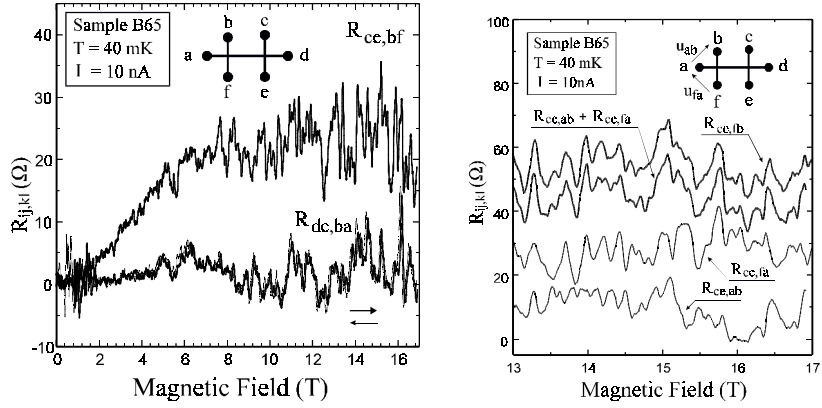

**Figure 2 left:** non-local magnetotransport data for 2 different contact configurations. The insert shows the contact scheme. **Right:** non-local data from different parts of the sample edge. The traces are shifted vertically for better visibility

Fig.2-left shows the data of two different non-local configurations $R_{ij,kl}$. $ij$ denotes the current contacts and $kl$ the voltage probes. In configuration $R_{ce,bf}$ the increase of the signal to an average value of about 20 Ω with increasing field indicates that edge state conduction is established at high fields. Because of the bulk conduction between the remote contacts b and f also a considerable amount of dissipation has to occur between these contacts. It is interesting to note that in this configuration there is a direct path along the edge from contact c to b and from e to f. Therefore the contact pair b - f seems to provide the first effective chance for EC's to be scattered to the bulk in order to produce dissipation. In the second configuration $R_{de,ba}$ there is always an "unused" contact between the current and remote voltage contacts along the sample edge. Consequently the EC's have the chance to be scattered to bulk before reaching the non-local voltage probe. Therefore most of the edge current is lost to the bulk before it reaches the voltage probes and therefore there is much less dissipation between the voltage probes in configuration $R_{de,ba}$. The signal below 2 Tesla is non reproducible noise which has no meaning for the interpretation. We have checked configuration $R_{ce,bf}$ again but measuring also the fluctuations between contacts f and a and contacts a and b separately (Fig.2-right). One can clearly see that the sum of $R_{ce,ab}$ and $R_{ce,fa}$ is equal to $R_{ce,fb}$ but the individual fluctuation patterns are significantly different. This strongly indicates that the individual patterns are generated independently. Remembering the general arguments given in section I, it is likely that the non-zero bulk resistance of the contact arms is responsible for the observed signals. This means that the fluctuation pattern can be generated by fluctuations of the bulk current in the voltage probes. As already explained, this bulk current in the contact arm is the returning current from a net edge current into the contact reservoir. If this edge current fluctuates because some EC's are bypassing the contact due to potential fluctuations, the observed signals can be understood to result from this bypass process. Consequently the actual observed patterns of Fig.2-right can be attributed to EC-backscattering in the contact arms b and f which generate their pattern independently. Turning back to Fig.2-left, the fluctuations in $R_{ce,bf}$ can be explained by EC's which are bypassing the contacts at the voltage probes. These bypassing EC's can enter at the next contact. This can be attributed to the signal of configuration $R_{de,ba}$ where the pattern is of comparable amplitude and may be generated by bypassing the "unused" contacts c and f.

## Acknowledgments

Financial support by Fonds zur Förderung der wissenschaftlichen Forschung (FWF) Vienna, Project: P10510 NAW.## References